\documentclass[twocolumn,epjc3]{svjour3}
\journalname{Eur. Phys. J. A}

\usepackage{graphicx}%
\usepackage{multirow}%
\usepackage{amsmath,amssymb,amsfonts}%
\usepackage{dsfont}%
\usepackage{mathrsfs}%
\usepackage{xcolor}%
\usepackage[title]{appendix}%
\usepackage[normalem]{ulem}

\usepackage[square,numbers,sort&compress]{natbib}
\allowdisplaybreaks

\begin{document}

\title{Chirally motivated $\pi\Sigma - \bar{K}N$ model in a finite volume}

\author{Peter C. Bruns\thanksref{e1} 
\and Ale\v{s} Ciepl\'{y}\thanksref{e2}}
\thankstext{e1}{e-mail: bruns@ujf.cas.cz}
\thankstext{e2}{e-mail: cieply@ujf.cas.cz}

\institute{Nuclear Physics Institute, Czech Academy of Sciences, 250 68 \v{R}e\v{z}, Czechia}

\date{\today}

\maketitle

\abstract{
We generalize the chirally motivated $\pi\Sigma - \bar{K}N$ coupled channels model to the cubic 
finite volume and use it to calculate the stationary energy spectrum that appears in a nice agreement 
with the spectrum obtained in the lattice QCD simulations by the BaSc collaboration. 
Several other comparisons with the BaSc results are made, in particular relating their pole positions 
of the meson-baryon scattering matrix to the two-pole picture of $\Lambda(1405)$ generated by our 
chiral model in the infinite volume.
}


\section{Introduction}
\label{sec:int}

The rigorous study of low-energy strangeness $S=-1$ meson-baryon (MB) scattering is hampered by the fact 
that neither perturbative QCD nor its low-energy effective field theory, Chiral Perturbation Theory, 
can be applied without model-dependent non-perturbative extensions in the energy region of interest. 
Such extensions are necessary due to the presence of an s-wave resonance just below the kaon-nucleon threshold, 
known as the $\Lambda(1405)$. Notably, almost all modern theoretical approaches which are based on interaction 
kernels derived from a chiral Lagrangian combined with coupled-channel methods \cite{Kaiser:1995eg,Oset:1997it,
Krippa:1998us,Garcia-Recio:2002yxy,Borasoy:2005ie,Cieply:2011nq,Ikeda:2012au,Guo:2012vv,Feijoo:2018den} come 
to the conclusion that there are actually {\it two poles} in the complex-energy plane near the $\bar{K}N$ threshold \cite{Oller:2000fj,Jido:2003cb,Hyodo:2011ur,Mai:2012dt,Cieply:2016jby,Bruns:2019bwg,Meissner:2020khl,Mai:2020ltx,Xie:2023cej}. 

A rather recent development is the application of lattice QCD (LQCD) to study $\pi\Sigma - \bar{K}N$ scattering 
\cite{BaSc:2023zvt,BaSc:2023ori}, while some work has already been done on the excited states seen in such reactions 
\cite{Takahashi:2009bu,Burch:2006cc,Engel:2012qp,Hall:2014uca}. There, QCD correlators are evaluated on a finite 
Euclidean space-time lattice employing Monte-Carlo methods to the QCD path integral. To connect the lattice results 
with the real world, one has to get under control the discretization errors, finite volume (FV) effects, 
and the extrapolation to physical quark masses. 
Resonance parameters like mass, width and other properties are not directly accessible 
from the correlators. Instead, some parametrization for the scattering amplitude is typically used to connect 
the results for the correlators with the pole position of the resonance on an unphysical Riemann sheet. 
For narrow isolated resonances, a simple Breit-Wigner parametrization might be sufficient, but for more complicated 
cases, sophisticated unitary models have to be employed, thereby introducing a model dependence to the analysis. 
Thus, from the perspective of the coupled-channel models, the lattice results constitute an additional kind 
of ``experimental data" to which the models can be fitted and compared. For the $\Lambda(1405)$, this has been 
done by several groups in the last few years \cite{MartinezTorres:2012yi, Molina:2015uqp,Pavao:2020zle}. 
To assess the inherent model dependence of the various extrapolations, it is desirable to consider a big variety 
of models, compare their outcome, and eventually find criteria for their quality. 

It is the main aim of the present contribution to demonstrate the consistency of the results from a recent LQCD simulation  
by the Baryon Scattering (BaSc) collaboration \cite{BaSc:2023zvt,BaSc:2023ori} with one particular chirally motivated 
coupled channel model, namely the Prague model developed and studied in \cite{Cieply:2011nq,Cieply:2009ea,Bruns:2021krp}. 
Specifically, we chose its most recent version \cite{Bruns:2021krp} as a representative example. The generalization 
of the model to the lattice setting of \cite{BaSc:2023ori} is quite straightforward, and will be explained only shortly 
in the next section. The discussion of the results, however, is given in some detail in Sec.~\ref{sec:res}. 
In particular, we study the movement of the $\Lambda(1405)$ poles when the volume is sent to infinity, and also 
(for the infinite volume case) the pole trajectories traversed upon varying the hadron masses and decay constants 
from their values in the lattice setting to the physical values. All in all, as we state in our summary in Sec.~\ref{sec:sum}, 
we find a satisfying agreement of the Prague model predictions with the lattice results of \cite{BaSc:2023ori}, 
and generate some further predictions which can be compared with future lattice simulations.

\section{Finite volume model formulation}
\label{sec:FVM}

In the Prague model \cite{Bruns:2021krp}, the s-wave meson-baryon scattering amplitude is of the form
\begin{equation}
  f_{0+}(s) = g(s)\lbrack 1-v_{0+}(s)G(s)\rbrack^{-1}v_{0+}(s)g(s)\,.
\label{eq:Pmodel}
\end{equation}
Here $v_{0+}$ is the s-wave projection of the interaction kernel as derived for the chiral MB Lagrangian, 
$G$ is a loop function describing the propagation of a MB pair, and the Yamaguchi form factors 
$g(q)=1/[1 + (q/\alpha)^2]$ provide the means to regularize the loop function integral with the parameters $\alpha$ 
representing regularization scales. In Eq.~(\ref{eq:Pmodel}) we use the notation $g(s)\equiv g(q(s))$, where $q(s)$ 
is the modulus of the center-of-mass (c.m.) momentum for the meson-baryon system for c.m. energy $\sqrt{s}$, 
see Eq.~(\ref{eq:qofs}) in Appendix \ref{sec:Ap2}. The building blocks of $f_{0+}$ are coupled-channel matrices, 
and its form is dictated by the strategy of re-summation of loop graphs with arbitrarily many insertions 
of the separable interaction kernel $v_{0+}$. This construction principle remains unchanged when restricting 
the model to a finite cubic lattice volume. Most importantly, the interaction kernel also remains unchanged, 
since it does not contain contributions from long-range exchanges (like e.g.~pion-exchange graphs), and we neglect 
exponentially suppressed FV corrections to masses and coupling constants. Strictly speaking, we should also modify 
the partial-wave projection, since rotational symmetry is broken on a cubic lattice. However, since the Prague model 
so far has a reliable parametrization only of the s-wave amplitude, such an attempt would be futile at this point. 
Thus, we have to restrict ourselves to the analysis of lattice states pertaining to irreducible re\-presentations 
of the cubic group which have a dominant overlap with infinite volume s-wave states. 

The general framework of quantum theories in a finite volume is well known \cite{Luscher:1985dn,Wiese:1988qy,Hasenfratz:1989pk,Luscher:1990ux,Luscher:1991cf}. Here, we follow the same strategy as e.g.~\cite{Doring:2011vk,MartinezTorres:2012yi, Molina:2015uqp, Severt:2020jzc} 
and modify only the loop function $G$ in our scattering amplitude Eq.~(\ref{eq:Pmodel}). In the infinite volume, 
the MB loop function employed in the Prague model \cite{Bruns:2021krp} is defined as
\begin{align}
 G_{\rm P}(q) &=  4\pi \!\int\! \frac{d^3l}{(2\pi)^3}\frac{[g(l)]^2}{l^2-q^2-{\rm i}\epsilon}   \nonumber \\
              &= [g(q)]^2 \left( \frac{\alpha^2 - q^2}{2\alpha} + {\rm i}\,q \right) \;,
\label{eq:Ginfv}
\end{align}
where $l^2\equiv|\vec{l}|^2$. Despite its simple non-relativistic appearance, our loop function has the correct 
relativistic absorptive part $\sim {\rm i}q$ due to the use of relativistic kinematics, see Eq.~(\ref{eq:qofs}) 
in Appendix \ref{sec:Ap2}. The form factors $g(l)$ regulate the propagation of particles with momenta $l \lesssim \alpha$ 
and have provided a convenient tool for extrapolations of the MB amplitudes off-the-energy shell since their introduction 
when the adopted coupled-channel model was designed in \cite{Kaiser:1995eg}. A comparison with an alternative of dimensionally 
regularized loop function was given e.g.~in \cite{Bruns:2022sio}, see the discussion following Eq.~(2.11) there.

We need to generalize Eq.~(\ref{eq:Ginfv}) to the case of a cubic FV $V=L^3$. For a free particle inside 
a finite cube of side length $L$, with periodic boundary conditions, the possible three-momenta are 
$\vec{p}_{n} = \frac{2\pi}{L}\vec{n}$, $\vec{n}\in\mathds{Z}^3$. The discretization of the integral 
in Eq.~(\ref{eq:Ginfv}) is straightforward and yields
\begin{equation}
G^{\rm FV}_{\mathrm{P}}(q;L) = \frac{4\pi}{L^3}\sum_{\vec{n}\in\mathds{Z}^3}
\frac{\left[g(p_{n})\right]^2}{p_n^2 - q^2} \;.
\label{eq:GPfv}
\end{equation}
Note that the summands depend only on $n^2:=|\vec{n}|^2$, and therefore we can just sum over this variable, 
with a weight factor $w(n^2)$ counting the multiplicity of lattice points $\vec{n}$ with a given 
$n^2$, $w(n^2) = 1,6,12,8,6 \ldots$ for $n^2 =0,1,2,3,4 \ldots$. 
Some additional details related to computation of the MB loop function are given in Appendix~\ref{sec:Ap1}.

Obviously, the formula Eq.~(\ref{eq:GPfv}) can be used as it is when analyzing the properties of the MB 
coupled channel system in the finite cubic volume. However, it should be noted that the series of poles present 
at the discrete momenta $q = \pm p_n$ replaces a branch cut inherent in the infinite volume loop function 
Eq.~(\ref{eq:Ginfv}) due to the multi-valued character of the momentum $q$ as a function of the Mandelstam variable $s$. 
Thus, there is no natural analogue of a second-sheet FV loop function via analytic continuation. However, one can 
devise expressions that approach the infinite volume physical sheet and second sheet loop functions 
for $L\rightarrow\infty$ everywhere except on the branch cut, see Appendix~\ref{sec:Ap2}. 

When evaluating the FV loop function at complex energies (and momenta) away from the real axis we found it 
most convenient to use the generalized formulas
\begin{align}
\label{eq:Lcorrections}
G^{\rm FV}_{\mathrm{P}}(q;L) &= [g(q)]^2 \!\left[ \tilde{G}_{\rm P}(q) \!+\! \Delta \tilde{G}_{\cal R}(q;L) \!+\! \Delta \tilde{G}_{\cal I}(q;L) \right], \nonumber \\
\Delta \tilde{G}_{\cal R}(q;L) &= \sum_{\vec{0}\not=\vec{n}\in\mathds{Z}^3} e^{-n\alpha L}\; ,  \\
\Delta \tilde{G}_{\cal I}(q;L) &= \sum_{\vec{0}\not=\vec{n}\in\mathds{Z}^3}\frac{1}{nL}\left[ e^{\pm {\rm i}qnL} -(1+n\alpha L)e^{-n\alpha L} \right]\; , \nonumber
\end{align}
where $n = |\vec{n}|$, the $\pm$ sign in the exponent on the last line matches the sign of the imaginary 
part of the momentum $q$, and the tilded $G$-functions stand for the loop function parts divided by 
the form-factor $g(q)$ squared, e.g.~$G_P(q) = [g(q)]^2 \tilde{G}_{\rm P}(q)$. The terms $\Delta \tilde{G}_{\cal R}(q;L)$ 
and $\Delta \tilde{G}_{\cal I}(q;L)$ represent FV corrections to the first ({\it pseudo-real}) and second ({\it phase-space}) 
parts of the infinite volume loop function, Eq.~(\ref{eq:Ginfv}), and apparently vanish in the $L \rightarrow \infty$ 
limit. We also note that, for a fixed $L$-value, the oscillations caused by the exp$(\pm {\rm i}qLn)$ term 
in $\Delta \tilde{G}_{\cal I}(q;L)$ are better suppressed at complex energies further from the real axis.

\section{Results and discussion}
\label{sec:res}

We start our analysis by looking at the hadron masses used by the BaSc collaboration, see Table III 
in their paper \cite{BaSc:2023ori}. While the kaon mass there, $M_K \approx 486$ MeV, is relatively close 
to the physical value, their pion mass $M_\pi \approx 204$ MeV still remains somewhat off, a common feature of present-day lattice simulations. However, we find it interesting that all BaSc hadron 
masses are very close to the trajectories of hadron masses adopted in \cite{Bruns:2021krp} when going 
to the SU(3) flavor symmetric limit, see the Appendix there. We demonstrate it in Fig.~\ref{fig:masses}, where 
the masses follow the paths from the SU(3) flavor symmetric point (for the scaling factor $x_{\rm SU3} = 0$) 
to their physical values (at $x_{\rm SU3} = 1$). As one can see, the BaSc hadron masses are very close 
to the depicted trajectories for $x_{\rm SU3} \approx 0.86$. The latter value can be viewed as 
an indication of how far are the LQCD predictions from the physical reality.

\begin{figure}[h]
\includegraphics[width=0.5\textwidth]{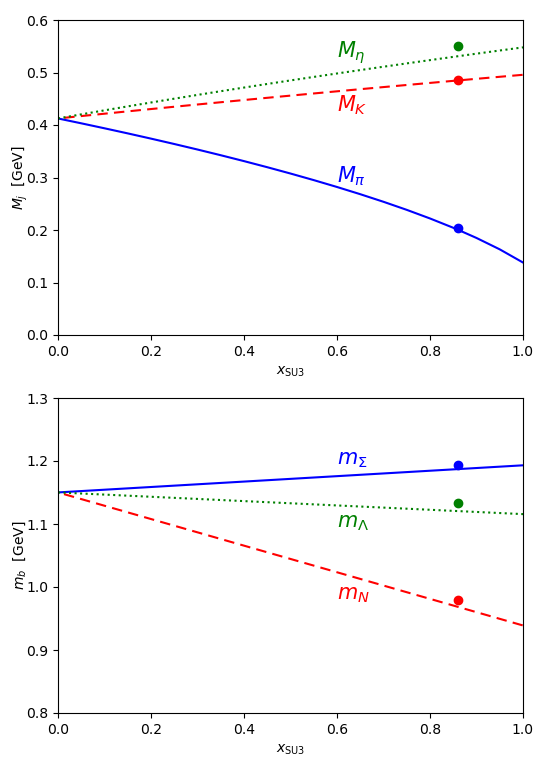}
\caption{The parametrization of hadron masses used in \cite{Bruns:2021krp} as a function of the $x_{\rm SU3}$ 
parameter that measures the amount of SU(3) flavor-symmetry breaking. The BaSc masses shown at $x_{\rm SU3} = 0.86$ 
are marked with the full dots. Top panel - meson masses, bottom panel - baryon masses.}
\label{fig:masses}    
\end{figure}

\begin{figure}[htb]
\includegraphics[width=0.5\textwidth]{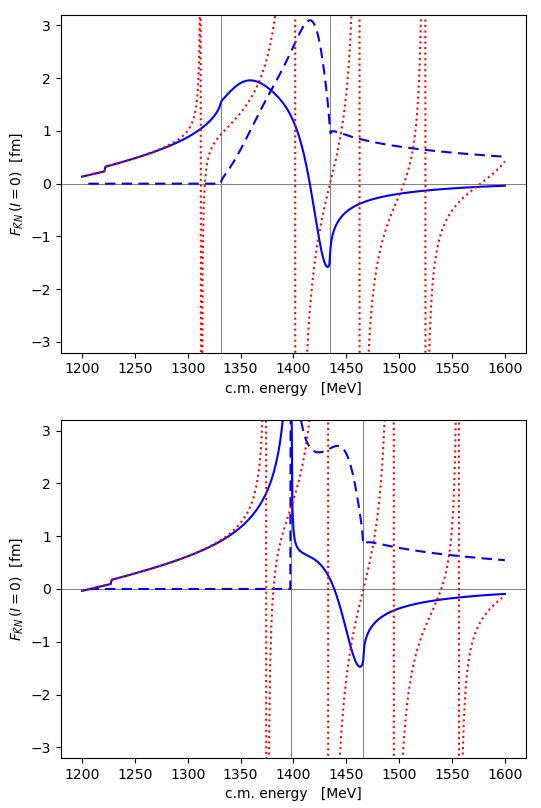}
\caption{The elastic isoscalar $\bar{K}N$ amplitude calculated in the finite volume for 
$L = 20$ GeV$^{-1}$ (dotted red lines) is compared with the real (continuous blue) and 
imaginary (dashed blue) parts of the amplitude generated by the Prague P0 model \cite{Bruns:2021krp}. 
The gray vertical lines mark the positions of the $\pi\Sigma$ and $\bar{K}N$ thresholds. 
Top panel - results for the physical hadron masses and meson decay constants, bottom panel - 
the same for the BaSc masses and decay constants.}
\label{fig:amplKN}    
\end{figure}

Let us follow this observation by a presentation and discussion of some results obtained with 
the chirally motivated $\pi\Sigma - \bar{K}N$ coupled channel model of Ref.~\cite{Bruns:2021krp} 
(also referred to as the P0 model in the following text)
re-formulated in the finite volume. If not specified otherwise, all our FV calculations were 
performed with $L=20$ GeV$^{-1}$, a value approximately matching the spatial extent $L = 4.051$ fm used 
in \cite{BaSc:2023ori}. In Fig.~\ref{fig:amplKN} we present the calculated elastic isoscalar 
$\bar{K}N$ amplitudes, plotted as a function of the c.m.~energy, for both the FV and infinite-volume 
formulations of the P0 model. Since the FV loop-function, Eq.~(\ref{eq:GPfv}), is manifestly real for real c.m. energies, 
so is the pertinent $\bar{K}N$ amplitude visualized by the red dotted lines. The real (continuous blue) 
and imaginary (dashed blue) parts of the amplitudes in the infinite volume are shown for comparison 
in the same Figure. In the top panel, the physical values of hadron masses and meson decay constants 
were used in the calculation, so the infinite volume curves correspond to those for the $K^{-}p$ amplitude 
presented in Fig.~2 in \cite{Bruns:2021krp}. In the bottom panel, the BaSc masses and meson decay 
constants were used and the channel thresholds are shifted accordingly. The unphysical hadron masses 
in this panel also lead to unphysical $\bar{K}N$ amplitudes including the one generated for the infinite 
volume. Though, it is worth noting the sharp variation of this amplitude around the (shifted) 
$\pi\Sigma$ threshold implying a nearby pole of the amplitude at the real axis or in its vicinity, 
a point we will discuss later on.

The FV amplitudes exhibit divergencies that relate to the poles realized at the points where 
$q^2 = p_n^2 = (2\pi/L)^{2}\,n^2$ when the MB loop function Eq.~(\ref{eq:GPfv}) 
is used instead of Eq.~(\ref{eq:Ginfv}) to generate the s-wave scattering amplitudes $f_{0+}$. 
We have kept in Fig.~\ref{fig:amplKN} the red dotted vertical lines that cross the real axis 
at points, where the amplitude values switch from $+\infty$ to $-\infty$. These points define 
the FV stationary energies that correspond to those reported in the LQCD calculations. 
In \cite{BaSc:2023ori}, the BaSc collaboration used the energy spectra of several irreducible 
lattice representations in fits of selected $K$-matrix coupled-channel models following a procedure 
outlined in \cite{Morningstar:2017spu} and utilized to make a connection to the real physical world. 
While their approach was fine to establish the existence of two poles of the scattering amplitude 
in the $\Lambda(1405)$ region, it was restricted to only two channels (the $\pi\Sigma$ and $\bar{K}N$ 
ones) and the adopted $K$-matrix models themselves are not as sophisticated as the modern coupled 
channels approaches based on interactions derived from the chiral Lagrangian. 

In principle, we could also tune the Prague model parameters to fit the BaSc energy spectra 
generated for the irreps shown in \cite{BaSc:2023ori} but it would be a more involved task. 
In the present work we just demonstrate how our finite-volume model does in comparison with 
the BaSc energy spectrum of the $G_{1u}(0)$ irrep that seems most relevant when one considers 
only the s-wave. In Table~\ref{tab:FVenerg} we present the stationary state energies generated by 
our FV model for two parameter settings, P0 and P2 taken from \cite{Bruns:2021krp, Cieply:2023saa}, 
and the BaSc energy spectrum of the $G_{1u}(0)$ irrep extracted from Fig.~8 in~\cite{BaSc:2023ori}. 
The P0 spectrum matches exactly the energies of divergencies observed in the bottom panel of Fig.~\ref{fig:amplKN} 
and the energy variations shown for the P0 and P2 spectra were obtained by varying the spatial extent 
by $\Delta L = 1$~GeV$^{-1}$, a value about 3 times larger than the standard deviation of the $L$-value 
adopted in \cite{BaSc:2023ori}. The two lowest energies (both below the $\bar{K}N$ threshold) appear 
particularly stable with respect to the $L$-variations in our model. It should be noted that the P2 model 
does not reproduce the $K^-p$ reactions data so well as the P0 model does due to complementing the model 
fit with the $\pi\Sigma$ photoproduction data. The P2 parameter setting is also closely related 
to a specific construction of the photoproduction kernel, see \cite{Cieply:2023saa} for details, and does 
not provide realistic predictions at higher energies, e.g.~for the $\Lambda(1670)$ pole position. 
Therefore, the P0 model results should be regarded more seriously while the P2 model spectrum 
of FV stationary energies was included in Table~\ref{tab:FVenerg} just to demonstrate a possible model dependence. 
As one can see, the correspondence between the P0 and BaSc energy spectra is remarkably good with only 
one energy being off by more than one standard deviation reported for the $G_{1u}(0)$ irrep. We would like 
to emphasize the point that these two energy spectra result from two very different approaches to QCD, 
one being the $\pi\Sigma - \bar{K}N$ coupled channel model based on chiral dynamics with parameters fitted 
to real physics data and the other the LQCD simulations of the MB interactions. 
It is really encouraging to see that so nice mutual agreement can be reached between these two 
distinctly different worlds.

\begin{table}[htb]
    \centering
    \begin{tabular}{cc|c}
       P0 energies & P2 energies & BaSc energies  \\  
      $ $[MeV]$ $  & $ $[MeV]$ $ &  $ $[MeV]$ $   \\ \hline
      $1375 \pm 2$ & $1371 \pm 2$ &  $1374 \pm 8$  \\
      $1432 \pm 1$ & $1429 \pm 1$ &  $1447 \pm 9$  \\
      $1494 \pm 6$ & $1509 \pm 8$ &  $1495 \pm 13$ \\
     $1556 \pm 12$ & $1594 \pm 16$ & $1566 \pm 12$   
    \end{tabular}
    \caption{A comparison of the finite volume stationary state energies generated by the P0 and P2 models 
    from \cite{Cieply:2023saa} with those obtained by the BaSc collaboration \cite{BaSc:2023ori} 
    for the $G_{1u}(0)$ irreducible representation.}
    \label{tab:FVenerg}
\end{table}

We now turn our attention to the poles of the scattering matrix. The Eq.~(\ref{eq:Lcorrections}) 
enables us to search for the poles of the FV amplitudes in the whole complex energy manifold. 
When performing the calculation with physical hadron masses and meson decay constants, for sufficiently 
large space volume (in the limit $L \rightarrow \infty$), we should be able to recover the pole positions 
generated in the original P0 model. We demonstrate this in Table \ref{tab:Lconverge}, where the pole 
positions are specified for increasing values of $L$. Since the typical values of the regularization scales 
are $\alpha \approx 0.3-0.7$ GeV, the finite volume correction $\Delta \tilde{G}_{\cal R}(q;L) \sim e^{-\alpha L}$ 
is quite small already for the lowest considered value $L=20$ GeV$^{-1}$. The other FV correction 
$\Delta \tilde{G}_{\cal I}(q;L)$ is more significant as its leading term in $L$ is proportional to 
$\exp (\pm {\rm i}qL)/L$ and it takes larger $L$-values to suppress it. As we also mentioned 
in Sec.~\ref{sec:FVM}, the Eq.~(\ref{eq:Lcorrections}) works better for complex energies further 
from the real axis which is reflected by the convergence of the two $\Lambda(1405)$ poles to their 
infinite volume positions. We have found that the $z_1$ pole reaches the latter for $L \approx 90$ 
(with the precision of $0.1$ MeV used in the Table) while $L \approx 190$ is needed for the $z_2$ pole.

\begin{table}[htb]
  \centering
  \begin{tabular}{c|c|c}
    $L$ [1/GeV] &  $z_1$ [MeV]    & $z_2$ [MeV]       \\ \hline 
      20        & (1365.8, -72.2) & (1569.8, -30.9)   \\
      40        & (1350.7, -46.4) & (1425.0, -24.5)   \\
      60        & (1352.0, -42.4) & (1426.2, -21.1)   \\
      80        & (1352.8, -42.6) & (1429.1, -26.1)   \\ 
     100        & (1352.7, -42.8) & (1429.5, -22.7)   \\ \hline 
      $\infty$  & (1352.7, -42.8) & (1428.5, -23.7)
  \end{tabular}
  \caption{The convergence of the generated pole positions to their infinite volume ones \cite{Bruns:2021krp}
           is demonstrated for increasing values of the lattice spatial extent $L$.}
  \label{tab:Lconverge}
\end{table}

\begin{figure}[htb]
\includegraphics[width=0.5\textwidth]{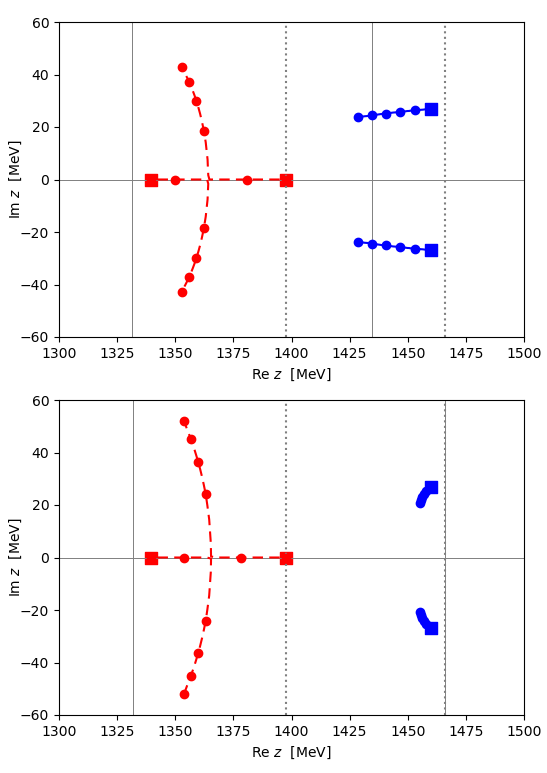}
\caption{The pole trajectories upon varying the hadron masses and meson decay constants 
from their lattice QCD values to those in the physical limit. The full squares show the pole positions 
for the former while the dots mark pole positions for the increasing scaling factor $x$ in steps 
of 0.2, the last point at $x=1$. Top panel - all hadron masses and meson decay constants are varied, 
bottom panel - only the pion mass is varied while all the other masses and decay constants are kept 
fixed at their lattice QCD values.}
\label{fig:poles}    
\end{figure}

Next, we would like to demonstrate how the pole positions found by the BaSc collaboration are 
related to those generated by the chirally motivated $\pi\Sigma - \bar{K}N$ coupled channel approaches. 
Once again, we make use of the P0 Prague model as a representative example of the latter and follow 
the movement of the poles when one varies the hadron masses and meson decay constants from their 
physical values to those adopted by the BaSc collaboration (and vice versa). In Fig.~\ref{fig:poles} 
we show the pole movements while the physical quantities are varied according to 
\begin{equation}
    \mathcal{Y}(x) = \mathcal{Y}_{\rm BaSc} + x\,(\mathcal{Y}_{\rm phys} - \mathcal{Y}_{\rm BaSc})\; ,
\label{eq:Xscale}    
\end{equation}
where $\mathcal{Y}$ represents the involved hadron masses and decay constants, and the linear scaling 
factor $x \in <0,\,1>$. The positions of the $\pi\Sigma$ and $\bar{K}N$ thresholds in the physical ($x=1$)
and BaSc ($x=0$) limits are visualized in the Figure by the continuous and dotted vertical lines, respectively.
While the top panel shows the movement of the poles when all hadron masses and meson decay constants 
are varied, the bottom panel demonstrates the impact of varying only the pion mass. Obviously, 
the $\bar{K}N$ threshold does not move in the latter case and the pertinent dotted and continuous 
lines coincide in the bottom panel.

The pole trajectories presented in the top panel of Fig.~\ref{fig:poles} clearly demonstrate that 
the pole positions found by the BaSc collaboration are related to those generated by the P0 model 
and it is natural to extrapolate that the same conclusion can be reached with the other chirally 
motivated coupled channel approaches. In particular, the virtual state reported close to the $\pi\Sigma$ 
threshold in \cite{BaSc:2023zvt, BaSc:2023ori} transforms into the lower mass $\Lambda(1405)$ pole 
when the hadron masses get to their physical values. We have checked that the variations of the meson 
decay constants have a negligible impact in this respect, most likely because their LQCD values used 
by the BaSc collaboration are already quite close to the physical ones. In fact, only the LQCD values 
of the pion and nucleon masses are significantly off their physical counterparts, which we have seen 
in Fig.~\ref{fig:masses} as well. The bottom panel of Fig.~\ref{fig:poles} shows that the non-physical 
pion mass used in the LQCD simulations is likely the main reason why a virtual state was found instead 
of a resonance in \cite{BaSc:2023zvt, BaSc:2023ori}, see also \cite{Xie:2023cej}. Indeed, as the pion mass is gradually restored 
to its physical value, the pole transforms into a proper resonant one. It is also worth noting that 
besides the pole close to the $\pi\Sigma$ threshold (the FV extension of the P0 model has it on the physical 
Riemann sheet, just marginally below the threshold) there is another pole further below the threshold at 
the $x=0$ limit. Its existence was not reported by the BaSc collaboration, but it is required by the analytical 
properties of the $\pi\Sigma - \bar{K}N$ coupled channel model. When one follows the movement of the two conjugate 
poles from their $x=1$ positions in Fig.~\ref{fig:poles}, they meet at the real axis (for $x \approx 0.28$ 
in the top panel) and then move along the real axis in opposite directions. Thus, there must be two poles 
in the $x=0$ limit.

\begin{figure}[htb]
\includegraphics[width=0.5\textwidth]{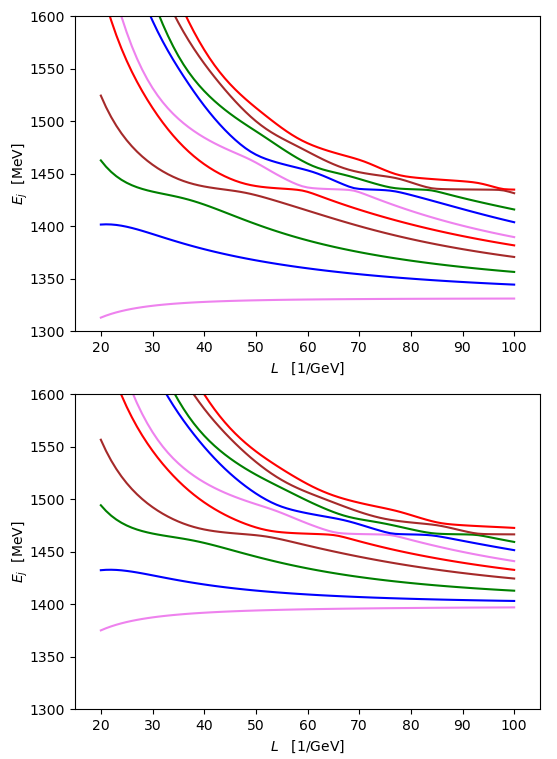}
\caption{The $L$-dependence of the finite volume energy spectra. Top panel - physical hadron masses and 
meson decay constants used to generate the ten lowest stationary state energies (Prague P0 model \cite{Bruns:2021krp}), 
bottom panel - BaSc hadron masses and meson decay constants used.}
\label{fig:FVspectra}    
\end{figure}

It is also well known that a presence of resonant states in FV simulations can be identified when one 
analyses the $L$-dependence of the stationary state energies \cite{Wiese:1988qy}. In Fig.~\ref{fig:FVspectra} 
we follow this idea utilizing the P0 model from \cite{Bruns:2021krp} and show the $L$-dependence of ten 
lowest energies. The top panel presents the results obtained with the physical hadron masses and meson 
decay constants while in the bottom panel the BaSc values were used for these quantities. The four 
lowest stationary energies at $L=20$~GeV$^{-1}$ match the positions of discontinuities observed 
for the FV $\bar{K}N$ amplitude in the respective top and bottom panels of Fig.~\ref{fig:amplKN}. 
According to \cite{Wiese:1988qy}, an appearance of the energy plateaus and the pertinent avoided 
energy levels crossings, that can be clearly noted in Fig.~\ref{fig:FVspectra}, is directly related 
to an existence of a resonance in the studied system. In the top panel, the plateau emerges at 
$E \approx 1435$~MeV, an energy only about $7$~MeV larger when compared with the infinite volume 
position of the (higher in mass) $z_2$ pole shown in Table \ref{tab:Lconverge}. The difference 
can be attributed to the coupled channels effects, in particular to the existence of the second 
$\Lambda(1405)$ pole generated by the chiral dynamics. On the other hand, the energy plateau 
at $E \approx 1460$~MeV observed in the bottom panel of Fig.~\ref{fig:FVspectra} almost exactly 
matches the position of the pole at $z = (1460,\; -27)$~MeV that our model generates for the BaSc 
setting in which the second pole degenerates into a state no longer recognized as another resonance.

\section{Summary}
\label{sec:sum}

We have presented a simple generalization of the chirally motivated $\pi\Sigma - \bar{K}N$ Prague model 
\cite{Bruns:2021krp} to the cubic finite volume. The FV formulation of the MB loop function enabled us to calculate 
the stationary energy levels of the coupled channel system and compare them with the appropriate $G_{1u}(0)$ irrep 
spectrum obtained recently by the BaSc collaboration \cite{BaSc:2023zvt, BaSc:2023ori}. Both spectra were 
found in remarkably good agreement, which provides a confidence for making more comparisons of our FV model 
with the LQCD simulations. We also find it interesting (and in a way encouraging) that the hadron masses 
used by the BaSc collaboration are quite close to the mass trajectories adopted in \cite{Bruns:2021krp} 
when going to the SU(3) flavor symmetric point.  The credibility of our FV model was also checked 
by confirming that the two poles of the scattering matrix found on the complex energy plane in the FV regime 
converge to the pole positions reported in \cite{Bruns:2021krp} when the cubic volume goes to infinity, 
i.e.~the spatial extent $L \rightarrow \infty$.

We have demonstrated that two poles of the scattering matrix found in the complex energy plane by the BaSc 
collaboration, a resonance just below the $\bar{K}N$ threshold and another one on the real axis close 
to the $\pi\Sigma$ threshold, transform into the standard two pole picture of $\Lambda(1405)$ when one 
reconciles gradually the LQCD hadron masses (and meson decay constants) with their physical values. 
We have also shown that the relatively large pion mass used in the LQCD simulations represents the main reason 
why one of the resonance poles (the one at lower mass and further from the real axis) degenerated into 
a virtual state reported by the BaSc collaboration. 

Finally, we have also looked at the $L$-dependence of the stationary energy spectra and shown that 
they exhibit features compatible with the discussed pole structure of the scattering matrix for 
both our FV version of the $\pi\Sigma - \bar{K}N$ Prague model as well as for the BaSc setting.

\section*{Acknowledgments}

We would like to thank B.~Cid-Mora for a helpful communication concerning the stationary energy spectra
generated by the BaSc collaboration.

\newpage
\onecolumn
\begin{appendices}
\section{Loop function in a finite volume}
\label{sec:Ap1}

When deriving the FV loop function $G^{\mathrm{FV}}_{\mathrm{P}}$ we found it convenient to separate it 
into a regularization-dependent {\it pseudo-real part} and a {\it phase-space part}. To this end, let us consider an integral 
representation for the phase-space part in the infinite volume, 
\begin{equation}
{\rm i}\,q = 4\pi\int\frac{d^3l}{(2\pi)^3} 
             \left[ \frac{1}{l^2-q^2-{\rm i}\epsilon} - \frac{l^2 + 3\alpha^2}{(l^2+\alpha^2)^2} \right]\; .
\label{eq:iqintegral}
\end{equation}
Going to a finite volume, we replace the integral on the r.h.s.~by a sum, multiply it with $[g(q)]^2$, 
and subtract the outcome from our result for $G^{\mathrm{FV}}_{\mathrm{P}}$ in Eq.~(\ref{eq:GPfv}):
\begin{equation}
  G^{\rm FV}_{\cal R}(q;L) := G^{\mathrm{FV}}_{\mathrm{P}}(q;L) 
    - [g(q)]^2 \,\frac{4\pi}{L^3} \sum_{\vec{n}\in\mathds{Z}^3} \left[ \frac{1}{p_n^2-q^2} - \frac{p_n^2 + 3\alpha^2}{(p_n^2+\alpha^2)^2} \right] 
    = [g(q)]^2 \,\frac{4\pi}{L^3} \sum_{\vec{n}\in\mathds{Z}^3} \frac{\alpha^2 - q^2}{(p_n^2+\alpha^2)^2}\;,
\label{eq:DeltaG}
\end{equation}
where the eligible discrete MB momenta are $\vec{p}_{n} = \frac{2\pi}{L}\vec{n}$. The poles in $q^2$ have 
dropped out in Eq.~(\ref{eq:DeltaG}), and the sum converges for all real $q^2$. 
The FV corrections in $G^{\rm FV}_{\cal R}$ are exponentially suppressed, as one can easily show employing the Poisson summation formula (PSF):
\begin{align}
  \tilde{G}^{\rm FV}_{\cal R}(q;L) &= G^{\rm FV}_{\cal R}(q;L) / [g(q)]^2 
     = 4\pi(\alpha^2-q^2) \int\frac{d^3 l}{(2\pi)^3} \left(\frac{2\pi}{L}\right)^3 
     \sum_{\vec{n}\in\mathds{Z}^3} \frac{\delta^3 (\vec{l}-\vec{p}_n )}{(l^2+\alpha^2)^2} \nonumber \\
  &\overset{\mathrm{PSF}}{=} 4\pi (\alpha^2-q^2) \int\frac{d^3 l}{(2\pi)^3}
     \sum_{\vec{k}\in\mathds{Z}^3} \frac{e^{{\rm i}L(\vec{k}\cdot\vec{l})}}{(l^2+\alpha^2)^2} 
     = \frac{\alpha^2-q^2}{2\alpha} \sum_{\vec{k}\in\mathds{Z}^3} e^{-\alpha L|\vec{k}|} \nonumber \\
  &= \frac{\alpha^2-q^2}{2\alpha} \left(1 + 6e^{-\alpha L} + 12e^{-\sqrt{2}\alpha L} + 8e^{-\sqrt{3}\alpha L} + \ldots\right)\;. 
\label{eq:ReGfv}
\end{align}
The first few terms of the expansion given in the last line yield a good approximation for $\alpha L \gtrsim 4$. 
Note that for real $q$ the first term in the expansion is equal to the real part of the infinite-volume loop function of Eq.~(\ref{eq:Ginfv}).

Formally, one can apply similar manipulations to the phase space ({\it pseudo-imaginary}) part:
\begin{align}
  \tilde{G}^{\rm FV}_{\cal I}(q;L) &= G^{\rm FV}_{\cal I}(q;L)  / [g(q)]^2 
    = 4\pi\int\frac{d^3l}{(2\pi)^3} \left(\frac{2\pi}{L}\right)^3 
    \sum_{\vec{n}\in\mathds{Z}^3} \delta^3 (\vec{l}-\vec{p}_n) \left[\frac{1}{l^2-q^2} - \frac{l^2 + 3\alpha^2}{(l^2+\alpha^2)^2}\right] \nonumber \\
  &\overset{\mathrm{PSF}}{=} ... \nonumber \\
  &= \pm {\rm i}\,q + \!\sum_{\vec{0}\not=\vec{k}\in\mathds{Z}^3}\frac{1}{|\vec{k}|L}\left(e^{\pm {\rm i}qL|\vec{k}|} 
     -(1+\alpha L|\vec{k}|)e^{-\alpha L|\vec{k}|}\right)\;,
\label{eq:ImGfv}
\end{align}
where the plus and minus signs in $\pm$ correspond to $\mathrm{Im}\,q>0$ and $\mathrm{Im}\,q<0$, respectively.
This expansion is not useful for real $q$, because it diverges at $q = \pm\, p_n$, and the oscillations 
of the exp$(\pm {\rm i}qL|\vec{k}|)$ terms are not damped. However, for complex $q$ away from the real axis, 
the first few terms yield a decent approximation for sufficiently large $L$. Note that there is some 
(exponentially suppressed) dependence on the real range parameter $\alpha$, which 
originates from the second term in Eq.~(\ref{eq:iqintegral}) and drops out in the infinite 
volume. 

\section{Second Riemann sheet treatment}
\label{sec:Ap2}

The loop function in the infinite volume, Eq.~(\ref{eq:Ginfv}), is usually considered as a function 
of the Mandelstam variable $s$ (the squared c.m.~energy), setting 
\begin{equation}
q\,\rightarrow\, q(s) :=\frac{1}{2\sqrt{s}}\sqrt{[s-(m_b+M_j)^2][s-(m_b-M_j)^2]}\; ,
\label{eq:qofs}
\end{equation}
where $m_b$ and $M_j$ denote the baryon and meson masses, respectively.
The square root should be defined in a way that it is positive for real $s>(m_b+M_j)^2$, and has 
a branch cut on the interval $((m_b+M_j)^2,\infty)$. We conventionally denote by $q$ the branch 
of the function that has a non-negative imaginary part. The branch cut structure is inherited by 
$G_{\mathrm{P}}(s)$, whose analytic continuation on the {\it second sheet}, traversing the branch cut 
from above, is then given by
\begin{equation}\label{eq:GinfvSheet2}
 G_{\mathrm{P}}^{\mathrm{II}}(s) = \frac{[\alpha-{\rm i}\,q(s)]^2}{2\alpha}[g(q(s))]^2\; ,
\end{equation}
i.e.~the sign of the ${\rm i}q$ term is reversed. Since in the FV expression Eq.~(\ref{eq:GPfv})
the branch cut is replaced by a series of poles, there is no natural analogue of a second-sheet function 
via analytic continuation. Fortunately, one can utilize the Eqs.~(\ref{eq:ReGfv} - \ref{eq:ImGfv}) 
and postulate
\begin{align}
  G^{\mathrm{FV(I)}}_{\mathrm{P}}(s;L)  &= G^{\rm FV}_{\cal R}(s;L) + G^{\rm FV}_{\cal I}(s;L)\; , \\
  G^{\mathrm{FV(II)}}_{\mathrm{P}}(s;L) &= G^{\rm FV}_{\cal R}(s;L) - G^{\rm FV}_{\cal I}(q;L)\; ,     
\label{eq:GFVsheet2}
\end{align}
with the plus sign (due to the chosen $q$ branch) taken in $G^{\rm FV}_{\cal I}$. 
As one can easily check, these expressions approach the infinite volume physical sheet and second sheet 
loop functions for $L\rightarrow\infty$ everywhere except on the branch cut. Alternatively, one could 
also define $q$ such that $\mathrm{Im}\,q<0$ in the lower complex $s$-plane, and use the $G^{\rm FV}_{\cal I}$ 
of Eq.~(\ref{eq:ImGfv}) with a negative sign in Eq.~(\ref{eq:GFVsheet2})\,.

\end{appendices}

\bibliography{2024EPJA}


\end{document}